\documentclass[aps,prl,twocolumn,groupedaddress]{revtex4}
\usepackage{graphicx}
\usepackage{amsmath}
\usepackage{relsize}
\usepackage{hyperref}
\usepackage{multirow}
\usepackage{color}
\usepackage{hhline}
\usepackage{float}
\restylefloat{table}

\begin{document}

\title{Questioning the cuprate paradigm - absence of superfluid density loss in several overdoped cuprates I}

\author{J. L. Tallon$^{1,\dag}$, J. G. Storey$^{1}$, J. W. Loram$^{2,\ddag}$, Jianlin Luo$^{3}$, C. Bernhard$^{4}$, I. Kokanovi\'{c}$^{2,5}$ and J. R. Cooper$^{2}$}

\affiliation{$^1$Robinson Research Institute, Victoria University of Wellington, P.O. Box 33436, New Zealand.}

\affiliation{$^2$Cavendish Laboratory, Cambridge University, Cambridge CB3 0HE, United Kingdom.}

\affiliation{$^3$Institute of Physics, Chinese Academy of Sciences, P.O.Box 603, Beijing, China.}

\affiliation{$^4$University of Fribourg, Physics Dept., Chemin du Musée 3, CH-1700 Fribourg, Switzerland.}

\affiliation{$^5$University of Zagreb, Department of Physics, P.O. Box 331, Zagreb, Croatia.}

\date{\today}
\begin{abstract}
{It is long established that overdoped cuprate superconductors experience a loss of superfluid density (SFD) with increasing doping, $p$, along with the decline in $T_c$. Such behavior is unconventional and suggests a depletion of the condensate by increasing pairbreaking or the growth of a second non-pairing channel. This led to a recent suggestion that the condensate arises from an {\it incoherent}  charge channel which progressively gives way  with overdoping to a second, {\it coherent} non-pairing channel. {\it Contra} these ideas, we report analysis of the field-dependent electronic specific heat of several cuprates from which we find no apparent loss of SFD with overdoping. The SFD per CuO$_2$ plaquette is found to rise progressively with overdoping from $p$ towards $(1+p)$, undiminished and much the same as the Hall number, thus implying that all available carriers contribute to the condensate. We suggest this could be the underlying intrinsic behavior for all cuprates. Our samples include (Y,Ca)Ba$_2$Cu$_3$O$_{7-\delta}$, Bi$_2$Sr$_2$CaCu$_2$O$_{8+\delta}$, La$_{2-x}$Sr$_x$CuO$_4$ and Tl$_2$Ba$_2$CuO$_6$, with the latter being the only exception. Our results signal a possible return to a more conventional picture.

}
\end{abstract}


\maketitle

\begin{figure}[h]
\centering
\includegraphics[width=60mm]{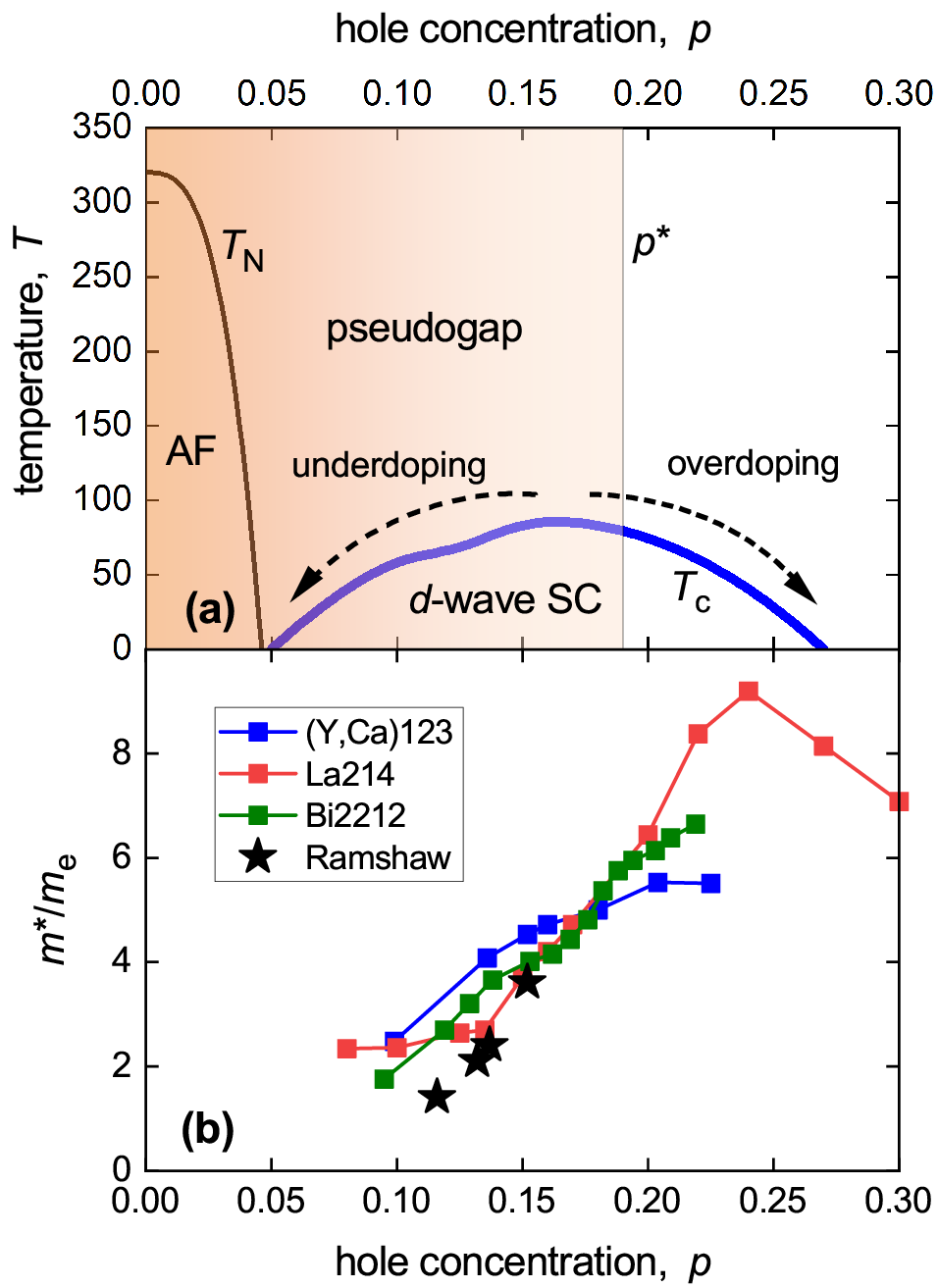}
\caption{\small
(a) Schematic phase diagram for cuprates as a function of doping, $p$. $p^*$ denotes critical doping and the region, $p<p^*$, is the pseudogap domain. Note the vertical boundary at $p^*$.
(b) electron mass enhancement deduced from specific heat as described in the text. }
\label{YCa123_1}
\end{figure}

The hole-doped cuprates exhibit a dome-shaped superconducting phase curve, rising to a maximum transition temperature, $T_{\textrm{c}}$, at {\it optimal} doping, $p$, then falling again, ultimately to zero, on the {\it overdoped} side (see Fig.~\ref{YCa123_1}(a) in End Matter). The {\it underdoped} region of the phase diagram is notable for the presence of a {\it pseudogap} domain, at present of uncertain origin, in which a gap opens on the Fermi surface around the ($\pi$,0) antinodes. This dramatically weakens the superconducting properties \cite{Loram2001}, most notably the superfluid density (SFD), the upper critical field and the critical current density \cite{BernhardAnom,Grissonanche,Talantsev}. As a consequence, 
there exists a sweet spot in the phase diagram, just past optimal doping, where the pseudogap closes and the superconducting properties are most robust. This is referred to here as {\it critical doping}, $p^*$, and it coincides with the location of maximal SFD. One long-established cuprate puzzle is the observation, initially from muon spin rotation ($\mu$SR) studies, that the SFD unexpectedly falls with overdoping beyond $p^*$ \cite{Boomerang,Niedermayer}. In the ground state of a conventional superconductor {\it all} electronic carriers condense into a coherent pairing state \cite{Waldram} so that increasing overdoping should see an increasing SFD. So the observed loss of SFD is unconventional and suggests a depletion of the condensate by increasing pairbreaking \cite{Lee-Hone,Kivelson} or the growth of a second non-pairing channel \cite{Culo2021}.

These original measurements \cite{Boomerang,Niedermayer} were conducted on overdoped Tl$_2$Ba$_2$CuO$_{6+\delta}$ (Tl2201). The $\mu$SR measurement allows extraction of the in-plane London penetration depth, $\lambda$ which is related to the SFD as follows
\begin{equation}
\lambda^{-2}  = \mu_0 e^2 (\rho_{\textrm{s}}/m^*) .
\label{SFD}
\end{equation}
Here $\mu_0$ is the permeability of free space, $e$ the electron charge, $m^*$ the effective electron mass and $\rho_{\textrm{s}}$ is the SFD, expected to be one half of the total carrier density \cite{Waldram}. 

Following these $\mu$SR studies came an independent study by Uemura {\it et al.} \cite{Uemuraboomerang} that confirmed the initial results in detail. 
Later $\mu$SR studies showed similar results for overdoped (Yb,Ca)Ba$_{0.6}$Sr$_{0.4}$Cu$_3$O$_{7-\delta}$ \cite{BoomerangYb}, and then (Y,Ca)Ba$_2$Cu$_3$O$_{7-\delta}$, Tl$_{0.5}$Pb$_{0.5}$Sr$_2$Ca$_{1-x}$Y$_x$Cu$_2$O$_7$ (Tl1212) \cite{BernhardAnom} and Ca$_x$La$_{1-x}$Ba$_{1.75-x}$La$_{0.25+x}$Cu$_3$O$_y$ (CLBLCO) \cite{Keren}. Some years later, a detailed study was reported on a large number of high-quality overdoped films of La$_{2-x}$Sr$_x$CuO$_4$ which showed a linear decline in $\lambda^{-2}$ with $T_{\textrm{c}}$ commencing abruptly from optimal doping \cite{Bozovic}. Such a decline is also seen in far-infrared optical studies \cite{Lemberger1,Lemberger2}. It would seem then that this collapse in SFD with increasing doping, hand in hand with the fall in $T_{\textrm{c}}$, is universal and canonical for overdoped cuprates.

This phenomenology was recently proposed by $\check{\textrm{C}}$ulo {\it et al.} \cite{Culo2021} to arise from the presence of two distinct charge sectors, one governed by {\it coherent} quasiparticle excitations (the non-pairing channel) and one by {\it incoherent} non-quasiparticle excitations characterized by Planckian dissipation (the pairing channel). In this view, the decrease in $n_{\textrm{s}}$ with increasing doping correlates with the loss of incoherent carriers, and the growth in density of coherent non-pairing quasiparticles complements the incoherent channel to complete a total of $(1+p)$ carriers per unit CuO$_2$ cell, consistent with the Luttinger sum rule for a large cuprate Fermi surface.

Here, we question this overdoped paradigm. From measurements of the electronic specific heat we find that the SFD does not collapse with overdoping, {\it at least over the doping range we have investigated down to around 50\% of maximum $T_{\textrm{c}}$}. Beyond that our conclusions must remain provisional. Our samples include Y$_{0.8}$Ca$_{0.2}$Ba$_2$Cu$_3$O$_{7-\delta}$ ((Y,Ca)123), Bi$_2$Sr$_2$CaCu$_2$O$_{8+\delta}$ (Bi2212) and La$_{2-x}$Sr$_x$CuO$_4$ (La214). In contrast, the same measurements in Tl$_2$Ba$_2$CuO$_{6+\delta}$ (Tl2201) support the accepted picture of rapid SFD collapse with overdoping. We are left with at least two major puzzles: (i) the apparent contradiction between the specific heat and the $\mu$SR, microwave and far-infrared measurements (at least on the overdoped side - they seem to concur for $p<p^*$), and (ii) the marked contrast between conventional behavior for Y123, Bi2212 and La214 on the one hand and, on the other, the unconventional SFD collapse for Tl2201 (and perhaps other cuprates). The inference is that the former cuprates are more `standard', while Tl2201 is an exception - despite the fact that it is particularly clean \cite{Bangura}, with an unusually low residual resistivity \cite{Lee-Hone}. A {\it Companion Paper} \cite{Tallon2025} contains much more detail.

\noindent {\bf The electronic specific heat.} Our specific heat measurements were done using a differential technique which allows separation of the electronic specific heat from the much larger phonon contribution, up to as high as 300 K or more \cite{Loram1993}. High precision measurements were made of the difference in specific heat between the sample and a closely related, but non-superconducting, reference sample produced e.g. by Zn or Co substitution. 
Furthermore, when field-dependent measurements are expressed as differences, for example $\Delta \gamma(H) = \gamma(H)-\gamma(0)$, any possible residual phonon specific heat is automatically eliminated.

The field-dependent specific heat may be integrated to obtain the shift in free energy as follows:


\begin{equation}
\Delta F(H) = F(H)-F(0) = \int_{T_{\textrm{f}}}^T T\Delta\gamma \text{ d}T - T\int_{T_{\textrm{f}}}^T \Delta\gamma \text{ d}T ,
\label{Fenergy}
\end{equation}
where $T_{\textrm{f}}$ is a temperature sufficiently above $T_{\textrm{c}}$ that superconducting fluctuations are absent. In Eq.~\ref{Fenergy} the first term is the magnetisation {\it internal energy}, while the second is the {\it entropy} term. Previously, this has been analyzed by means of the London equation which has an $H\ln(H/H_{\textrm{c2}})$ scaling, where $H_{\textrm{c2}}$ is the upper critical field \cite{Campbell,Naqib}. However, our samples are polycrystalline and we show in Appendix A (EndMatter) how the London equation may be integrated around all field angles relative to the basal plane of a grain. This gives
\begin{equation}
\Delta F(H) = - \left(\frac{1}{2}\right) \frac{a\phi_0 H V_{\textrm{ga}}}{8\pi\lambda^2} \ln\left(\frac{H}{e^2bH_\textrm{c2}}\right) ,
\label{FHaoPoly}
\end{equation}
\noindent where $\phi_0$ is the flux quantum for pairs and $V_{\textrm{ga}}$ is the volume per gram atom. Thus, both $\lambda^{-2}$ and $H_{\textrm{c2}}$ may be extracted by fitting the experimental data for $\Delta F(H)$.

A complication arises in that $\lambda$, appearing in Eq.~\ref{FHaoPoly} is itself field dependent for a $d$-wave superconductor where the local approximation breaks down near the nodes of the order parameter. The $H$-dependence of $\lambda^{-2}$ is calculated by Amin, Franz and Affleck \cite{Amin}. We have fitted their calculations using the functional form $\eta = \lambda^{-2}(H)/\lambda^{-2}(0) = \exp(-B^{\alpha}/\beta)$, where $B=\mu_0H$ in units of tesla, and the $T$-dependent coefficients $\alpha$ and $\beta$ are given in Appendix B. These coefficients cover the full temperature range, but at $T=0$ the $H$-dependence is more accurately given by $\alpha = 0.44$ and $\beta = 6.7$. Here we extract $\lambda^{-2}$ from the $\Delta F$ data by noting that $H/H_{\textrm{c2}}$ (or $B/B_{\textrm{c2}}$) inside the logarithm of Eq.~\ref{FHaoPoly} can be expressed as $W/\Delta F_{\textrm{ns}}$ where $\Delta F_{\textrm{ns}}$ is the condensation free energy, $(1/2\mu_0)V_{\textrm{ga}}B_{\textrm{c}}^2$, and $W = B B_{\textrm{c1}} V_{\textrm{ga}}/(2 \mu_0 \ln\kappa)$. Here $B_{\textrm{c1}}$ is the lower critical field and $B_{\textrm{c}}$ is the thermodynamic critical field. Thus Eq.~\ref{FHaoPoly} can be rearranged to read:
\begin{equation}
\Delta F(H,T) = - \left(\frac{a}{2}\right) W \ln\left(\frac{1}{e^2b} \frac{W}{\Delta F_{\textrm{ns}}}\right) .
\label{FHaoPolyRe}
\end{equation}
For any value of $H$ the $T$-dependent value of $W$ can be solved self-consistently using the experimental $\Delta F(H,T)$ and $\Delta F_{\textrm{ns}}(T)$. From $W$ it is straightforward to extract $B_{\textrm{c1}}(H)$ and hence $\lambda^{-2}(H)$ and then use Amin, Franz and Affleck to convert to $\lambda^{-2}(0)$. Effectively, the value of $H_{\textrm{c2}}$ is determined here from $H_{\textrm{c}}^2/H_{\textrm{c1}}$. We now apply these equations to the analysis of $\Delta F$.

\noindent {\bf Y$_{0.8}$Ca$_{0.2}$Ba$_2$Cu$_3$O$_{7-\delta}$.} In all, seven doping states for the one sample were measured: three underdoped starting at $T_{\textrm{c}} = 55$ K, one at optimal doping ($T_{\textrm{c}} = 85$ K), and three overdoped, the last with $p = 0.225$ ($T_{\textrm{c}} = 54$ K). Measurements were done in zero-field and in a field of 13 tesla, under a field-cooled protocol \cite{Luo}. 
Fig.~\ref{YCa123_2}(a) in Appendix C shows what is effectively the raw data for $\Delta\gamma(13,T)$ being the difference in raw data for $\gamma(13,T)$ and $\gamma(0,T)$. Fig.~\ref{YCa123_2}(b) shows the entropy, $\Delta S(13,T)$, obtained by integrating $\Delta \gamma(13,T)$. Below 14 K, $\gamma(H,T)$ shows a small field-independent upturn, the origin of which is not understood. This is largely absent in the difference $\Delta\gamma(H,T)$, and conveniently, we ignore this by simply extrapolating $\Delta S(13,T)$ to zero at $T=0$, as it must do. Irrespective of this, and despite the obvious broadening of the transition seen in $\Delta\gamma(13,T)$ with overdoping, one can see that the entropic weight in the transition is preserved across the four overdoped samples, certainly showing no sign of collapse. Fig.~\ref{YCa123_2}(c) shows the free energy $\Delta F(13,T)$ obtained by integrating $\Delta S(13,T)$ and again one can see that the free energy remains high in the overdoped region but falls a little. This small drop is largely due to the noted increasing broadening and is not due to a fall in SFD, as we will now see.

\begin{figure}
\centering
\includegraphics[width=60mm]{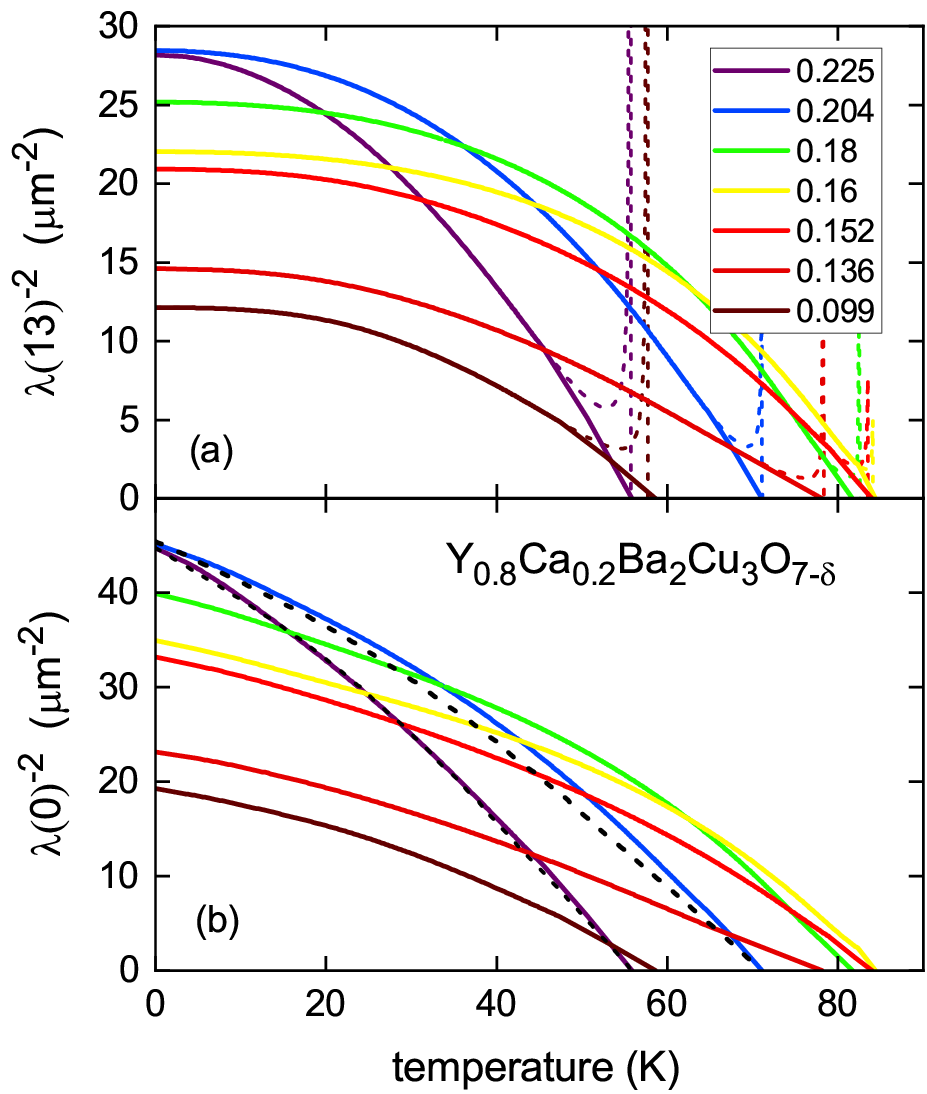}
\caption{\small
(a) The penetration depth at 13 tesla calculated from $\Delta F(13,T) = F(13,T)-F(0,T)$ shown in Fig.~\ref{YCa123_2}(c) for Y$_{0.8}$Ca$_{0.2}$Ba$_2$Cu$_3$O$_{7-\delta}$ using the London model (Eq.~\ref{FHaoPoly}). The dashed divergences seen near $T_{\textrm{c}}$ occur because $H/H_{\textrm{c2}}$ diverges as $T\rightarrow T_{\textrm{c}}$. We extrapolate each curve from above this breakdown to smoothly approach zero at $T_{\textrm{c}}$. (b) The penetration depth at zero field calculated from $\lambda(13)^{-2}$ in panel (a) using the renormalization due to Amin et al. \cite{Amin} given by Eq.~\ref{AminFit}. }
\label{YCa123_3}
\end{figure}

Fig.~\ref{YCa123_3} shows (a) the superfluid density $\lambda(13,T)^{-2}$ calculated using Eq.~\ref{FHaoPolyRe}, and (b) the zero-field superfluid density $\lambda(0,T)^{-2}$ deduced using the renormalization due to Amin et al. \cite{Amin} given by Eq.~\ref{AminFit}. In panel (a) the calculation breaks down on approaching $T_{\textrm{c}}$ where $H/H_{\textrm{c2}}$ begins to diverge. This is evident in the dashed curves. Therefore, because $\lambda(13,T)^{-2}$ must vanish at $T_{\textrm{c}}$ we have simply smoothly extrapolated each curve to zero at $T_{\textrm{c}}$ and this is shown by the solid curves in Fig.~\ref{YCa123_3}(a). These curves flatten at low $T$ due to the presence of the applied field. Renormalising the solid curves to $H=0$ gives the solid curves in Fig.~\ref{YCa123_3}(b) and these now show the expected linear-in-$T$ dependence at low $T$ and an overall $T$-dependence very similar to that expected for a weak-coupling $d$-wave superconductor - see black dashed curves for the two most overdoped samples. For the most overdoped sample there is almost exact agreement.

\begin{figure}
\centering
\includegraphics[width=60mm]{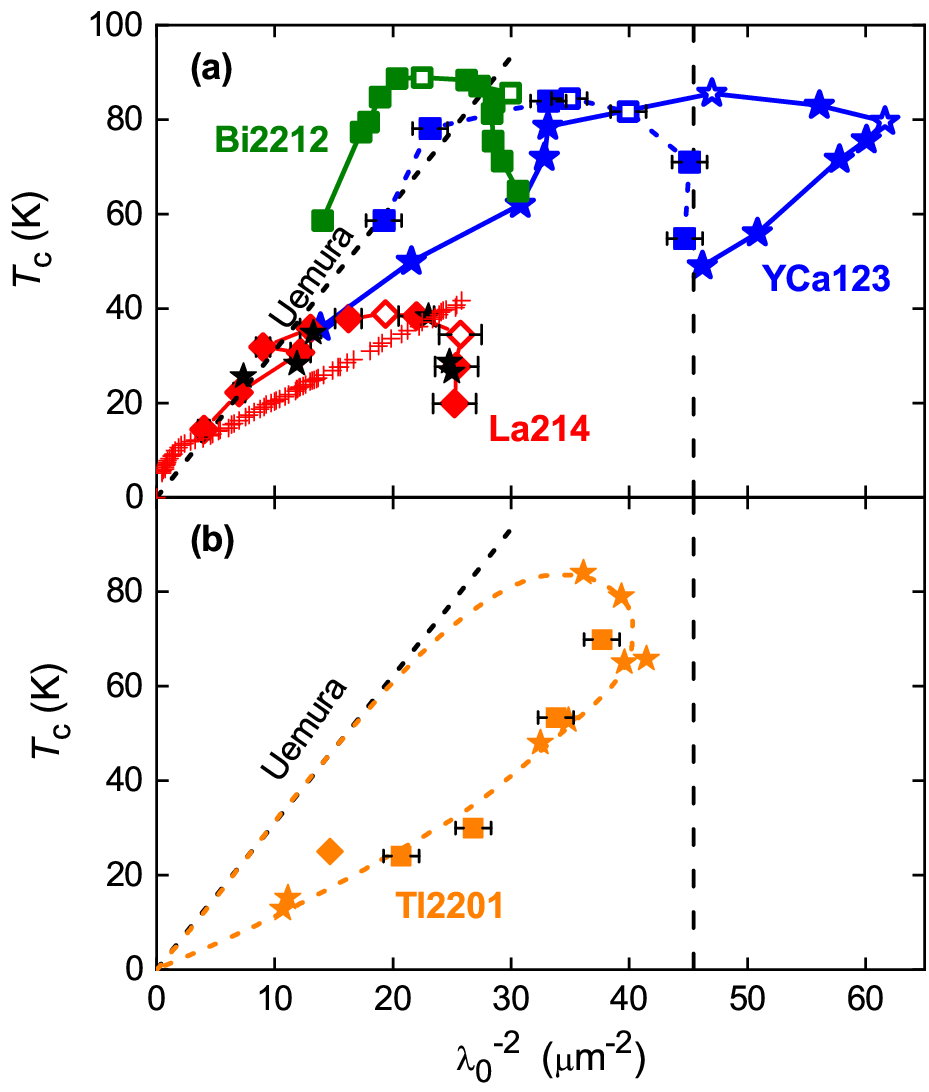}
\caption{\small
(a) Uemura plot, $T_{\textrm{c}}(p)$ vs $\lambda_0(p)^{-2}$, for our results for (Y,Ca)123 (blue squares), Bi2212 (olive squares), La214 (red diamonds) \cite{Paradigm} and (Y,Ca)123 obtained from $\mu$SR \cite{BernhardAnom} (blue stars). Red crosses show the data of Bo\v{z}ovi\'{c} {\it et al.} for La214 films \cite{Bozovic}. (b) $\lambda_0^{-2}$ for Tl2201: this work (orange squares) and previous $\mu$SR results \cite{Niedermayer,Uemuraboomerang} (orange stars) or microwave data (orange diamond). }
\label{YCa123_4}
\end{figure}

The key result of Fig.~\ref{YCa123_3} is that, for the three overdoped samples, $\lambda(0,T)^{-2}$ does not fall with overdoping but continues to rise then plateaus in the most heavily overdoped sample with $T_{\textrm{c}} = 54$ K. This contrasts the $\mu$SR results for this system \cite{Boomerang,Niedermayer} which are summarized in Fig.~\ref{YCa123_4}(a) in the form of a Uemura plot \cite{Uemura} of $T_{\textrm{c}}$ versus $\lambda(0,0)^{-2}$ (blue squares). They are compared with other cuprates \cite{Paradigm} to be discussed below. In each case the open data point at highest $T_{\textrm{c}}$ represents optimal doping and the next open data point represents critical doping, $p\approx0.19$. Error bars are shown in the figure and discussed elsewhere \cite{Tallon2025}.

\begin{figure}
\centering
\includegraphics[width=60mm]{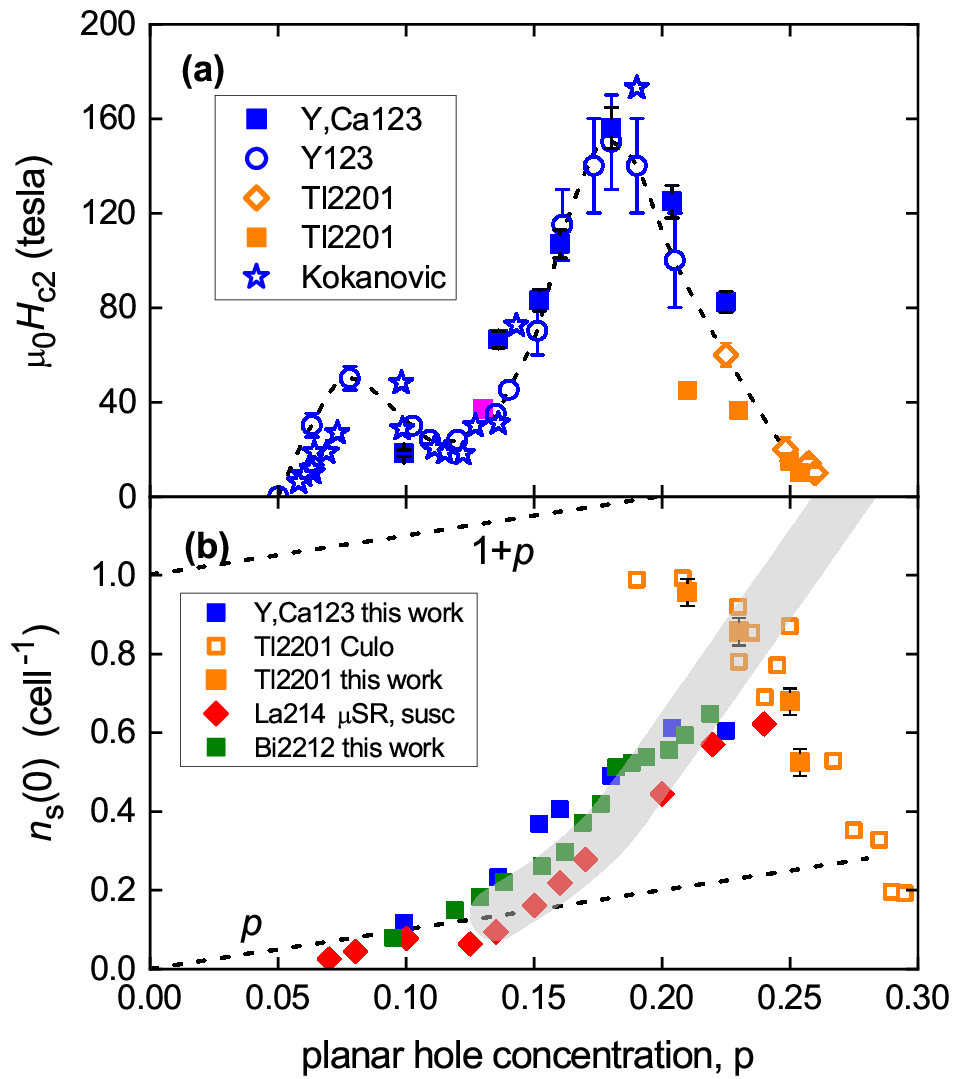}
\caption{\small
(a) $H_{\textrm{c2}}(T=0)$ calculated using the London model for (Y,Ca)123 (blue squares) along with values reported in ref. \cite{Grissonanche} (open blue circles, open orange diamonds) and by Kokanovi\'{c} and Cooper \cite{Kokanovic} (open blue stars).
(b) $n_{\textrm{s}}(0)$ vs $p$ inferred from $\lambda_0^{-2}$: (Y,Ca)123 (blue squares), La214 (red diamonds) and Bi2212 (olive squares). The shaded strip summarises the carrier filling, $n_{\textrm{n}}$, for Bi2201 \cite{Putzke}, Tl2201 \cite{Tam} and La214 \cite{Legros2022} which matches our rising $n_{\textrm{s}}(0)$ values. In contrast, values of $n_{\textrm{s}}(0)$ for Tl2201 calculated from our measured $\gamma(H,T)$ (orange squares) and as reported by $\check{\textrm{C}}$ulo {\it et al.} \cite{Culo2021} (open orange squares) decline steadily with doping. }
\label{ns0}
\end{figure}

From these calculations we also obtain values of the upper critical field, $H_{\textrm{c2}}(T=0,p)$, and these are plotted as a function of doping, $p$, in Fig.~\ref{ns0}(a) as shown by the blue filled squares. Also shown are the $p$-dependent values of $H_{\textrm{c2}}$ reported by Grissonanche {\it et al.} \cite{Grissonanche} from high-field thermal conductivity (open blue circles) and by Kokanovi\'{c} and Cooper \cite{Kokanovic} (blue stars). The general agreement is excellent across the phase diagram.

\noindent {\bf La$_{2-x}$Sr$_{x}$CuO$_4$ and Bi$_2$Sr$_2$CaCu$_2$O$_{8+\delta}$.} Unfortunately we have almost no data for the field-dependent specific heat of La214, however we do have $\mu$SR and ac susceptibility measurements of $\lambda^{-2}$ on the same set of samples on which our zero-field specific heat measurements were made \cite{Panagopoulos}. $T=0$ values are plotted in Fig.~\ref{YCa123_4}(a) by the red diamonds with, again, open symbols denoting optimal and critical doping. These values are in excellent agreement with those reported by Uemura from $\mu$SR \cite{Uemuraboomerang} (black stars). Notably, there is no evidence of a collapse in SFD with overdoping - entirely consistent with our observation, for the same sample batch, that the overdoped $\gamma_{\textrm{res}}=0$ and that the jump in specific heat at $T_{\textrm{c}}$ does not diminish \cite{Tallon2025}. This contradicts the data for La214 films reported by Bo\v{z}ovi\'{c} {\it et al.} \cite{Bozovic}, reproduced by the red crosses in Fig.~\ref{YCa123_4}(a), which shows an immediate linear collapse of SFD beyond optimal doping while we report a significant increase.

Previously \cite{Paradigm}, we have analysed the field-dependent specific heat of Bi2212 \cite{Loram2001} using the Fetter and Hohenberg  parameterisation of the London equation \cite{Fetter}. As for (Y,Ca)123 and La214, it revealed a rather constant value of $\lambda_0^{-2}$ across the accessible overdoped region, again, consistent with the sustained value of the jump, $\Delta \gamma_{\textrm{c}}(0)$, and the finding that $\gamma_{\textrm{res}}=0$ \cite{Tallon2025}. For consistency, we have reanalysed the raw $\Delta\gamma(H,T)$ data for Bi2212 \cite{Tallon2025} using the present Hao and Clem coefficients for the London equation \cite{HaoClem}. The new derived values of $\lambda_0^{-2}$ are shown in Fig.~\ref{YCa123_4}(a) by the olive green squares. The absolute values have risen by about 12\% with the new analysis but the key result still holds: $\lambda_0^{-2}$ remains undiminished with overdoping, and even rises a little. In summary, our specific heat data for these three materials, (Y,Ca)123, La214 and Bi2212, show no sign of a diminishing SFD with overdoping and they rather emphatically contrast the behavior of Tl2201 to which we now turn.

\begin{figure}[H]
\centering
\includegraphics[width=60mm]{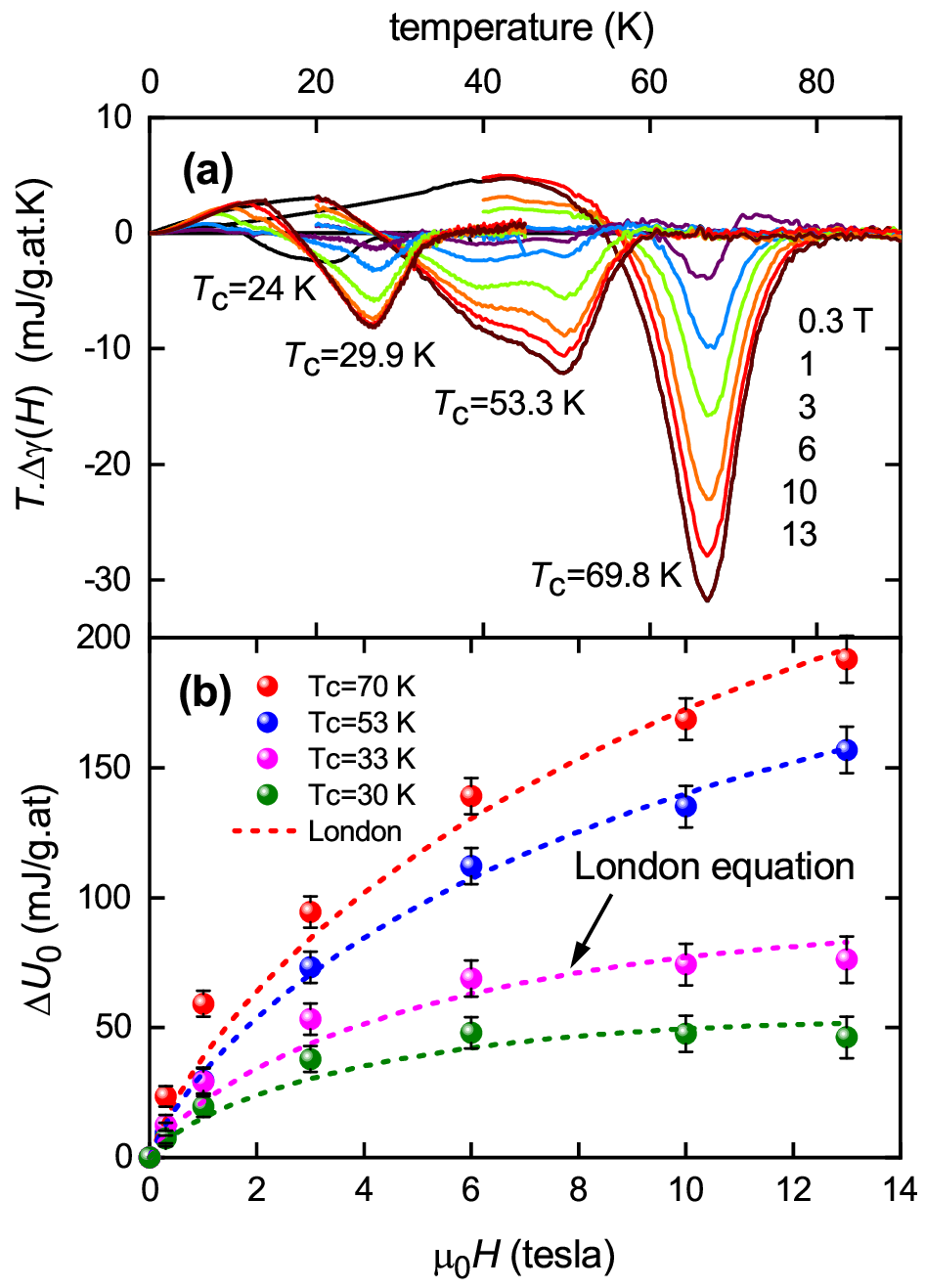}
\caption{\small
(a) The $T$-dependence of $T \Delta \gamma(H)$ for four progressively overdoped samples of Tl2201 with $T_{\textrm{c}} \approx$ 70 K, 53 K, 30 K and 24 K. Some lower-$T$ data is omitted to avoid cluttering. The field strength for each run is shown by the legend. (b) The field-dependence of the ground-state internal energy $\Delta U_0$ for each overdoped sample of Tl2201 obtained by integrating the data in (a) using Eg.~\ref{Fenergy}. The dashed curves are the London equation using Eg.~\ref{FHaoPoly}. The inferred parameters $\lambda(0)$ and $B_{\textrm{c2}}$ are plotted in Fig.~\ref{YCa123_4}(b) and Fig.~\ref{ns0}(a), respectively. }
\label{Tl2201_1}
\end{figure}

\noindent {\bf Tl$_2$Ba$_2$CuO$_{6+\delta}$.} 
Fig.~\ref{Tl2201_1}(a) shows the measured raw data for $T\Delta \gamma(H,T)$ for four overdoped samples with $T_{\textrm{c}}$ = 70 K, 53 K, 30 K and 24 K. Some of this data has already been reported elsewhere \cite{Radcliffe} but not subjected to a London analysis. The second of these samples is evidently inhomogeneous, showing a broadened double transition. To avoid cluttering, the low-$T$ data is omitted except for the single black curves for 13 tesla. For the 24 K sample only the 13 tesla curve is shown. The area under these curves is the magnetisation {\it internal} energy and one can see immediately that the weight of the condensate is progressively, and heavily, depleted.

The integrated energy, $\Delta U(H,0)$, is plotted versus field in Fig.~\ref{Tl2201_1}(b) and fitted to the London model by the dashed curves. The fits are reasonable given that the two most overdoped samples have reached the limits of the Hao-Clem parameterisation. The obtained values for $\lambda_0^{-2}$ are plotted in Fig.~\ref{YCa123_4}(b) by the orange squares, together with the data reported earlier from $\mu$SR studies \cite{Niedermayer,Uemuraboomerang} (orange stars) and from microwave studies \cite{Broun} (orange diamond). The agreement is excellent. The orange dashed curve is a guide to the eye that links the three data sets, and all three show a sustained loss of SFD across the overdoped region. This contrasts our data for (Y,Ca)123, Bi2212 and La214, and signals a different behavior for Tl2201 amongst the cuprates we have studied.


\medskip
\noindent {\bf Discussion.} Here and elsewhere \cite{Tallon2025}, we have presented three model-free diagnostics that indicate that the SFD in (Y,Ca)123, Bi2212 and La214 does not shrink with overdoping: $\gamma_{\textrm{res}}$ remains zero, and both the anomaly height $\Delta\gamma(T_{\textrm{c}})$ and the magnetization free energy, $\Delta F(H)$ remain undiminished. We may add a fourth diagnostic. For both (Y,Ca)123 and Bi2212 it has been shown that, while the zero-field condensation energy, $U_0$, falls with overdoping, the BCS ratio, $U_0/(\gamma_{\textrm{n}}T_{\textrm{c}}^2)$ remains constant \cite{Loram2001}. 
This immediately signals retention of the full condensate weight down to ~50\% of maximum $T_{\textrm{c}}$.  Taking $U_0 = (1/2)\mu_0H_{\textrm{c}}^2$ and $H_{\textrm{c}} = \phi_0/(2\sqrt{2}\mu_0\pi\xi\lambda)$  we find: 
\begin{equation}
U_0/(\gamma_{\textrm{n}}T_{\textrm{c}}^2) = \frac{1}{\mu_0 \gamma_{\textrm{n}}} \left(\frac{\phi_0}{4\pi\xi T_{\textrm{c}} }\right)^2 \lambda^{-2} .
\label{U0}
\end{equation}
\noindent Noting that $2\Delta_0/(k_{\textrm{B}}T_{\textrm{c}})$ remains constant with overdoping \cite{TallonFluc,ShenIncoherent} then it follows from the relation $\xi_0 = \hbar v_{\textrm{F}}/(\pi\Delta_0)$ that if $U_0/(\gamma_{\textrm{n}}T_{\textrm{c}}^2)$ remains constant then so will $\lambda_0^{-2}$. This applies within a multiplicative factor of $\gamma_{\textrm{n}} v_{\textrm{F}}^2$ which itself changes little with overdoping. Thus, irrespective of models and detailed quantification we conclude just from the zero-field specific heat that $\lambda_0^{-2}$ does not collapse with overdoping.

That stated, our results can be quantified using the approach of $\check{\textrm{C}}$ulo {\it et al.} \cite{Culo2021}. They identified an incoherent channel from which the condensate arises, and a non-pairing coherent channel which grows with overdoping while the pairing channel depletes. The number, $n_{\textrm{s}}$, of superfluid carriers per CuO$_2$ plaquette (of volume $V_{\textrm{cell}}$) is obtained by inverting Eq.~\ref{SFD}:
\begin{equation}
n_{\textrm{s}}(0) = \lambda_{\textrm{L}}^{-2}(0) \frac{m_{\textrm{e}}}{\mu_0 e^2} V_{\textrm{cell}} \left(\frac{m^*}{m_{\textrm{e}}} \right) .
\label{SFcell}
\end{equation}

\noindent We follow these authors in estimating the mass enhancement, $(m^*/m_{\textrm{e}})$, from the specific heat for a 2D metal:
\begin{equation}
\gamma_{\textrm{n}} = \frac{\pi k_{\textrm{B}}^2 N_{\textrm{A}} a^2 m_{\textrm{e}}}{3 \hbar^2} \left(\frac{m^*}{m_{\textrm{e}}} \right) .
\label{g2D}
\end{equation}

For (Y,Ca)123, $\gamma(T)$ contains contributions from the double CuO$_2$ planes and the CuO$_{1-\delta}$ chains. As shown in our Companion Paper \cite{Tallon2025}, by comparing with several other cuprates we estimate that each CuO$_2$ plane contributes $7.30 \pm 0.29$ mJ/mol/K$^2$ (with an implicit chain contribution of 40\% of the total $\gamma_{\textrm{n}}$ at high-$T$, almost independent of oxygen deficiency). This gives a sense of the absolute accuracy for our measurement of $\gamma(T)$ (while precision is more than a factor of 10 higher). From this planar contribution we use Eq.~\ref{g2D} to obtain a value of $m^*/m_{\textrm{e}} = 4.93 \pm 0.19$. This is a little larger than the values reported by Ramshaw {\it et al.} \cite{Ramshaw} which are shown by stars in Fig.~\ref{YCa123_1}(b). Those authors interpreted the rising value of $m^*$ in terms of the approach to a quantum critical point. We expect, however, that the increase in $m^*(p)$ is likely due to the closing of the pseudogap. Either way, it is the low-$T$ value of $\gamma(T)$ which is pertinent to $m^*$, not the $p$-independent, high-$T$ value where the double-peaked Fermi window for $\gamma$ lies outside of the low-energy DOS features, whether a diverging peak (vHs) or pseudogap \cite{Tallon124}. What is the appropriate low-$T$ value of $\gamma$ to use? Here we adopt the value of $S(T)/T$ at $T_{\textrm{c}}$ which is identical to the average value, $\langle\gamma_{\textrm{n}}\rangle$, of $\gamma_{\textrm{n}}(T)$ in the range $0 \leq T\leq T_{\textrm{c}}$. 
Fig.~\ref{YCa123_1}(b) shows $m^*/m_{\textrm{e}}$ for (Y,Ca)123 calculated in this way. The value increases across the underdoped region as the peudogap closes, and then remains largely unchanged for $p>0.19$.

From these values of $m^*/m_{\textrm{e}}$ and our reported values of $\lambda^{-2}(0,0)$ we calculate the doping dependence of $n_{\textrm{s}}(0)$ using Eq.~\ref{SFcell} and this is plotted by the blue squares in Fig.~\ref{ns0}(b). 
The overall trend with doping clearly suggests a crossover in $n_{\textrm{s}}(0,p)$ from $p$ towards $(1+p)$. If this is the incoherent channel as suggested by $\check{\textrm{C}}$ulo {\it et al.} \cite{Culo2021}, {\it it grows, rather than declines, with overdoping}, at least for $p\leq0.24$. In Fig.~\ref{ns0}(b) we plot their results for $n_{\textrm{s}}(0,p)$ for Tl2201 by the open orange squares and add our own values deduced here from $\Delta\gamma(H)$ (solid orange squares). There is good agreement and the collective data evolves in opposite fashion to that for (Y,Ca)123.

Turning to La214 and Bi2212, we calculate $m^*/m_{\textrm{e}}$ in the same way using $\langle\gamma_{\textrm{n}}\rangle$ and these are plotted vs $p$ in Fig.~\ref{YCa123_1}(b). The values of $n_{\textrm{s}}(0)$ thus calculated from $\lambda^{-2}$ are plotted in Fig.~\ref{ns0}(b) by the full red diamonds and olive squares, respectively. Despite the very different Uemura plots for (Y,Ca)123, La214 and Bi2212, by normalising the SFD to its value per CuO$_2$ plaquette, they seem to come together into a single common behavior, progressing from $p$ towards $(1+p)$, and contrasting the picture of $\check{\textrm{C}}$ulo {\it et al.} \cite{Culo2021}. The broad grey strip in Fig.~\ref{ns0}(b) combines the $p$-dependence of the Hall number, $n_{\textrm{H}}$, measured for Bi$_2$Sr$_2$CuO$_{6+\delta}$ (Bi2201) \cite{Putzke} and Tl2201 \cite{Tam} with the normal-state filling number, $n_{\textrm{n}}$, for La214 from THz cyclotron studies \cite{Legros2022}. This strip trends much the same as our $n_{\textrm{s}}(0)$ values, thus suggesting that all {\it available} carriers contribute to the condensate over the doping range considered. 

Our results contradict the $\mu$SR data on (Y,Ca)123, shown in Fig.~\ref{YCa123_4}(a) (blue stars), which reveal a marked fall in SFD with overdoping, as do Tl1212 \cite{BernhardAnom} and CLBLCO \cite{Keren}. For most of our overdoped (Y,Ca)123 there is a clear excess in SFD found in the $\mu$SR studies. This remains a puzzle that needs to be resolved either in favor of the $\mu$SR or specific heat data, or perhaps both. We do note that the angular average in Eq.~\ref{MHaoPoly}, End Matter, is probably more complex for orthorhombic Y123 materials \cite{HaoClem2}. But it is the {\it combination} of the four model-free diagnostics: $\gamma_{\textrm{res}}$, the jump $\Delta\gamma/\gamma_{\textrm{n}}$, $\Delta F(H)$, and $U_0/(\gamma_{\textrm{n}}T_{\textrm{c}}^2)$ that is most compelling in revealing an SFD that remains undepleted with overdoping.

Finally, we note that specific heat studies on overdoped La214 by Wang {\it et al.}  \cite{Wen214} showed a substantial, and growing, $\gamma_{\textrm{res}}$ which is completely absent in our samples. This would surely indicate some degree of sample preparation dependence for overdoped samples. It seems that, with the closure of the pseudogap at $p^*$, 
we begin to see these various non-generic variations appearing. Certainly the rapid, monotonic collapse in overdoped SFD observed by Bo\v{z}ovi\'{c} {\it et al.} \cite{Bozovic}, is contradicted by our data and could arise from Josephson coupling in their films combined with a falling gap amplitude, $\Delta_0(p)$ \cite{TallonFluc,ShenIncoherent}.

\medskip
\noindent {\bf Conclusions.} From the electronic specific heat we find there is no loss of condensate weight with overdoping in three canonical cuprates, (Y,Ca)123, La214 and Bi2212 for the range of $p$ studied, down to 50\% of maximum $T_{\textrm{c}}$. The only exception is Tl2201, which remains puzzling. 
Our finding of non-universal behavior in the pairing condensate across the overdoped region (i.e. in the absence of the pseudogap) is an important new puzzle in cuprate physics. Further, the stark discrepancy between our specific heat data for La214 and that of Wang {\it et al.} \cite{Wen214} suggests a material-preparation dependence. If so, then there remains the possibility that Tl2201 might be prepared in such a way that it, too, does not display overdoped SFD loss. Perhaps even, this is the canonical {\it intrinsic} behavior for all cuprates, contrary to the current widely-accepted paradigm.

\medskip
The authors wish to thank N.E. Hussey and A. Carrington for helpful critique.

\medskip
$^\dag$ jeff.tallon@vuw.ac.nz

$^\ddag$ deceased November 2017.

\clearpage

\noindent {\bf END MATTER}

\noindent {\bf Appendix A - the London Equation.} Hao and Clem \cite{HaoClem} proposed the following form for the London equation for the field-dependent magnetisation

\begin{equation}
\mu_0 M = \frac{a\phi_0 V_{\textrm{ga}}}{8\pi\lambda^2} \ln\left(\frac{H}{bH_\textrm{c2}}\right) ,
\label{Hao}
\end{equation}
with $\mu_0$ and $\phi_0$ being the permeability of free space and the flux quantum respectively, and $a = 0.77$ and $b = 1.44$ are constants, applicable for the range $0.02<H/H_{\textrm{c2}}<0.3$. We have converted here to SI units. $V_{\textrm{ga}}$ is the volume per {\it gram atom} and is related to the molar volume, $V_{\textrm{M}}$, by $V_{\textrm{ga}} = (1/N_{\textrm{f}}) V_{\textrm{M}}$ where $N_{\textrm{f}}$ is the number of atoms per formula unit. Although Eq.~\ref{Hao} is derived using standard mean-field Ginzburg-Landau theory, the force between 3D or 2D (pancake) vortices goes as $\ln(H)$ \cite{deGennes} so it may have more general validity. Eq.~\ref{Hao} may be integrated to give the associated free energy function:

\begin{equation}
\Delta F = -\frac{a\phi_0 H V_{\textrm{ga}}}{8\pi\lambda^2} \ln\left(\frac{H}{ebH_\textrm{c2}}\right) .
\label{GHao}
\end{equation}

For a polycrystalline sample of randomly-oriented grains we must replace $H$ in eq.~\ref{Hao} by $H\cos\theta$ and $H_{\textrm{c2}}$ by $H_{\textrm{c2}}(\theta)$ and integrate around the angular sphere, where $\theta$ is the angle between the applied field and the perpendicular to the basal plane. For tetragonal symmetry we have \cite{Gray}
\begin{equation}
H_{\textrm{c2}}(\theta) = H_{\textrm{c2}}\left[\cos^2\theta + \gamma_{\textrm{an}}^{-2} \sin^2\theta\right]^{-1/2}  ,
\label{Hc2}
\end{equation}
where $\gamma_{an}=m_c^*/m_{ab}^*$ is the anisotropy parameter. Because the condensation energy ($\propto H_{\textrm{c1}}$$\cdot$$H_{\textrm{c2}}$) must be independent of field orientation, we also have:
\begin{equation}
H_{\textrm{c1}}(\theta) = H_{\textrm{c1}}\left[\cos^2\theta + \gamma_{\textrm{an}}^{-2} \sin^2\theta\right]^{+1/2}  ,
\label{Hc1}
\end{equation}
where $H_{\textrm{c2}}$ and $H_{\textrm{c1}}$ are the values for field normal to the basal plane ($\theta = 0$).

Since for all cuprates $\gamma_{\textrm{an}}^{-2} \gg 1$, we may approximate $H_{\textrm{c2}}(\theta)$ by $H_{\textrm{c2}} /\cos\theta$ and $H_{\textrm{c1}}(\theta)$ by $H_{\textrm{c1}}\cos\theta$. Inserting in eq.~\ref{Hao} and averaging:
\begin{equation}
\mu_0 \langle M\rangle = \frac{a\phi_0 V_{\textrm{ga}}}{8\pi\lambda^2} \int_0^{\pi/2} \cos\theta\ln\left(\frac{H\cos^2\theta}{bH_\textrm{c2}}\right) \sin\theta d\theta,
\label{MHaoPoly}
\end{equation}
This may be integrated exactly to:
\begin{equation}
\mu_0 \langle M\rangle = -\left(\frac{1}{2}\right) \left(\frac{a\phi_0}{8\pi\lambda^2}\right) V_{\textrm{ga}} \ln\left(\frac{H}{ebH_\textrm{c2}}\right) ,
\label{MHaoPolyInt}
\end{equation}
and, as before, integrating with respect to $H$ gives the free energy
\begin{equation}
\Delta F_{\textrm{poly}} = -\left(\frac{1}{2}\right) \frac{a\phi_0 H V_{\textrm{ga}}}{8\pi\lambda^2} \ln\left(\frac{H}{e^2bH_\textrm{c2}}\right) .
\label{GHaoPoly}
\end{equation}
The effect, therefore, of a polycrystalline random distribution of grain orientation is that Eq.~\ref{GHao} is preserved with $H_{\textrm{c1}}$ renormalised to $\frac{1}{2}H_{\textrm{c1}}$ and $H_{\textrm{c2}}$ renormalised to $e^{-1} H_{\textrm{c2}}$.

\bigskip
\noindent {\bf Appendix B - field dependence of SFD.} 
Amin, Franz and Affleck \cite{Amin} have shown that for $d$-wave symmetry $\lambda$ is $H$-dependent, and at $\mu_0H$ = 13 tesla, $\lambda^{-2}(H) = 0.613 \lambda^{-2}(0)$, a rather large effect in the field range of interest here. Because $\lambda$ is a function of field the integrals above are not strictly valid. The problem can be resolved as follows. We have found that the results of Amin {\it et al.} can be modelled using:
\begin{equation}
\lambda^{-2}(H,T) = \lambda^{-2}(0,T) \exp(-H^{\alpha}/\beta) ,
\label{AminFit}
\end{equation}
where
\begin{equation}
\alpha = 0.4421 + 1.3625x + 0.5095x^2 -10.222x^3 ,
\end{equation}
\begin{equation}
\beta = 6.8657 + 30.809x + 362.048x^2 -751.111x^3 ,
\end{equation}
with $x = T/T_{\textrm{c}}$. In our companion paper we show how these equations can be used to solve the full problem. In the meantime Eqs.[14-16] may be used to renormalise $\lambda(H,T)^{-2}$ to $\lambda(0,T)^{-2}$.

\bigskip
\noindent {\bf Appendix C - supporting figures.}

\begin{figure}[H]
\centering
\includegraphics[width=60mm]{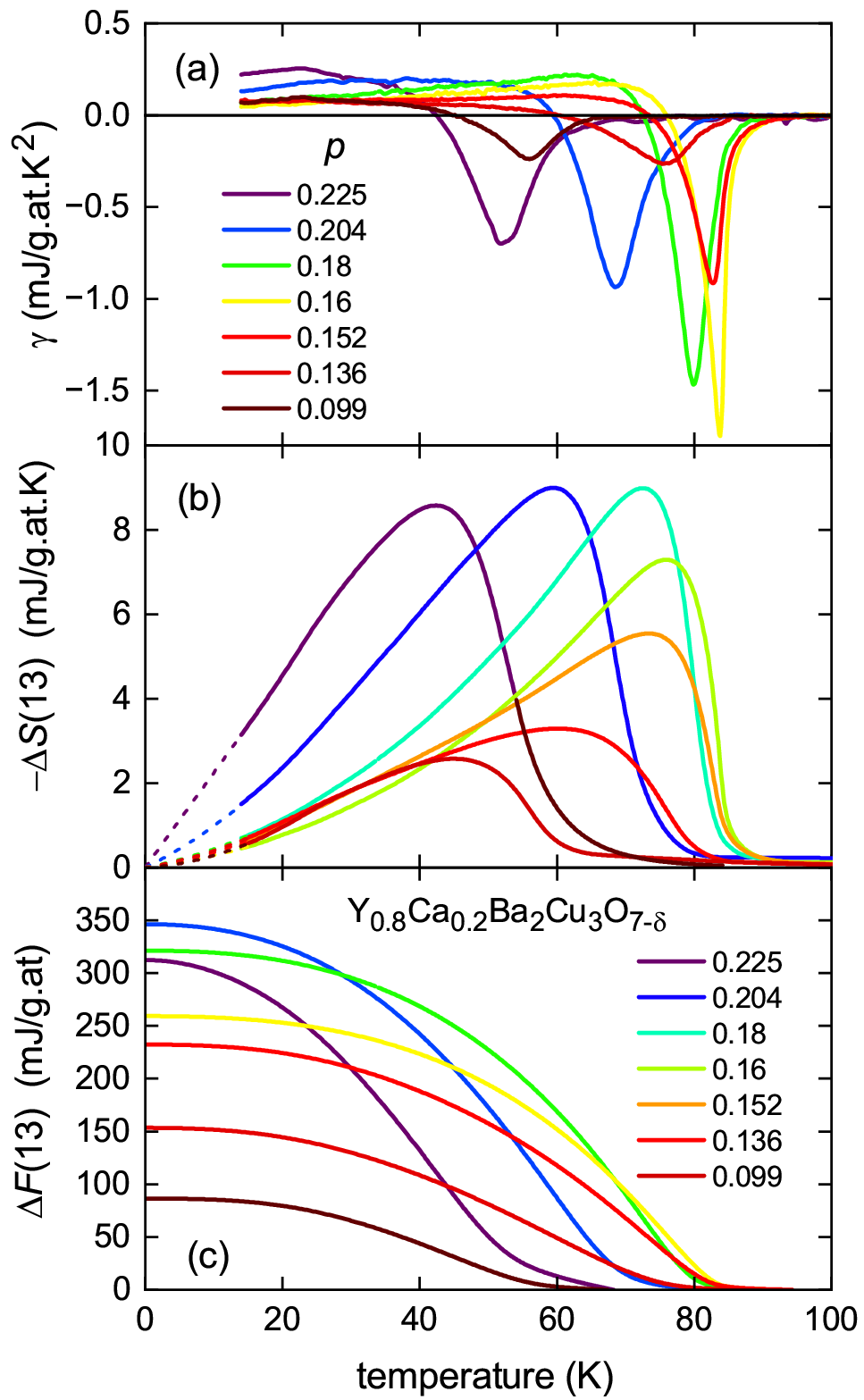}
\caption{\small
(a) The change in $\gamma$ with field, $\Delta\gamma(13,T) = \gamma(13,T)-\gamma(0,T)$, for Y$_{0.8}$Ca$_{0.2}$Ba$_2$Cu$_3$O$_{7-\delta}$. The transition is very sharp around optimal doping, but broadens with overdoping probably due to a random distribution of Ca pairs \cite{Naqib}; (b) the magnetic entropy, $\Delta S(13,T) = S(13,T)-S(0,T)$, obtained by integrating $\Delta\gamma(13,T)$; and (c) the change in magnetic free energy $\Delta F(13)$ obtained by integrating $\Delta S(13)$. The curves cover seven doping states as annotated, one optimal, three underdoped and three overdoped. }
\label{YCa123_2}
\end{figure}

\end{document}